\def\year{2021}\relax
\documentclass[letterpaper]{article} %
\usepackage{aaai21}  %
\usepackage{times}  %
\usepackage{helvet} %
\usepackage{courier}  %
\usepackage[hyphens]{url}  %
\usepackage{graphicx} %
\usepackage{natbib}  %
\usepackage{caption} %
\usepackage{xcolor}
\usepackage{balance}

\usepackage{xspace}

\newcommand{\new}[1]{#1}%

\newcommand{\hide}[1]{}
\newcommand{\xhdr}[1]{\vspace{1.5mm}\noindent{{\bf #1.}}} %

\newcommand{\etc}{\textit{etc.}\xspace}
\newcommand{\eg}{\textit{e.g.}\xspace}

\newcommand{\etal}{\textit{et al.}\xspace}
\newcommand{\fig}{Fig.\xspace}
\newcommand{\sect}{\S}
\newcommand{\reddit}{reddit\xspace}

\newcommand{\papersite}{\scriptsize{\texttt{https://behavioral-data.github.io/news\_labeling\_reddit/}}}

\usepackage{amsmath}
\newcommand{\E}{\mathrm{E}}
\newcommand{\var}{\mathrm{Var}}

\usepackage{enumitem}
\setlist{nosep}

\graphicspath{{figs/}}

\title{Political Bias and Factualness in News Sharing\\across more than 100,000 Online Communities}

\author{
Galen Weld,\textsuperscript{1} Maria Glenski,\textsuperscript{2} Tim Althoff\textsuperscript{1} \\
\normalsize{\textsuperscript{1}{Paul G. Allen School of Computer Science and Engineering}} \\
\normalsize{\textsuperscript{2}{National Security Directorate, Pacific Northwest National Laboratory}}
}

\begin{document}

\nocopyright

\maketitle

\begin{abstract}
As civil discourse increasingly takes place online, misinformation and the polarization of news shared in online communities have become ever more relevant concerns with real world harms across our society. %
Studying online news sharing at scale is challenging due to the massive volume of content which is shared by millions of users across thousands of communities.
Therefore, existing research has largely focused on specific communities or specific interventions, such as bans.
However, understanding the prevalence and spread of misinformation and polarization more broadly, across thousands of online communities, is critical for the development of governance strategies, interventions, and community design.
Here, we conduct the largest study of news sharing on \reddit to date, analyzing more than 550 million links spanning 4 years. We use non-partisan news source ratings from Media Bias/Fact Check to annotate links to news sources with their political bias and factualness.
We find that, compared to left-leaning communities, right-leaning communities have 105\% more variance in the political bias of their news sources, and more links to relatively-more biased sources, on average. 
We observe that \reddit users' voting and re-sharing behaviors generally decrease the visibility of extremely biased and low factual content, which receives 20\% fewer upvotes and 30\% fewer exposures from crossposts than more neutral or more factual content. This suggests that \reddit is more resilient to low factual content than Twitter. 
We show that extremely biased and low factual content is very concentrated, with 99\% of such content being shared in only 0.5\% of communities, giving credence to the recent strategy of community-wide bans and quarantines.
\end{abstract}
\section{Introduction}\label{sec:intro}
Biased and inaccurate news shared online are major concerns that have risen to the forefront of public discourse regarding social media in recent years. Two thirds of Americans get at least some of their news content from social media, but less than half expect this content to be accurate~\cite{shearer_2018}.
Globally, only 22\% of survey respondents trust the news in social media ``most of the time'' \cite{reuters_news_report_2020}.
Internet platforms such as Twitter, Facebook, and \reddit account for an ever-increasing share of the dissemination and discussion of news \cite{geiger_2019_online_news}.  %

Harms caused by biased and false news have substantial impact across our society.
Polarized content on Twitter and Facebook has been shown to play a role in the outcome of elections \cite{Recuero2020HyperpartisanshipDA, Kharratzadeh2017USPE}; and misinformation related to COVID-19 has been found to have a negative impact on public health responses to the pandemic~\cite{Tasnim2020ImpactOR, Kouzy2020CoronavirusGV}.
Developing methods for reducing these harms requires a broad understanding of the political bias and factualness of news content shared online, but studying news sharing is challenging for three reasons: (1) the scale is immense, with billions of news links shared annually, (2) it is difficult to automatically quantify bias and factualness at scales where human labeling is often infeasible \cite{Rajadesingan2020QuickCL}, and (3) the distribution of links is complex, with these links shared by many millions of users and thousands of communities.

While previous research has led to important insights on specific aspects of news sharing, such as user engagement \cite{Risch2020TopCO}, fact checking \cite{vosoughi2018spread,choi2020rumor}, specific communities \cite{Rajadesingan2020QuickCL}, and specific rumors \cite{vosoughi2017rumor,qazvinian2011rumor}, large scale studies of news sharing are critical to understanding polarization and misinformation more broadly, and can inform community design, governance, and moderation interventions.

In this work, we present the largest study to date of news sharing behavior on \reddit, one of the most popular social media websites. 
We analyze all 559 million links submitted to \reddit from 2015-2019\footnote{August 2019 was the most recent month of data available at the time of this study.}, including 35 million news links submitted by 1.3 million users to 135 thousand communities. We rate the bias and factualness of linked-to news sources using Media Bias/Fact Check (MBFC),\footnote{While bias and factualness may vary from story to story, news source-level ratings maximize the number of links that can be rated, and are commonly used in research \cite{bozarth2020higher}.} which considers how news sources favor different sides of the left-right political spectrum (bias), and the veracity of claims made in specific news stories (factualness)  (\sect\ref{sec:method}).

In our analyses, we examine: the \textit{diversity of news within communities} (\sect\ref{sec:communities_not_equal}), and how this diversity is composed of both the differences between community members and individual members' diversity of submissions; the \textit{impact of current curation and amplification behaviors} on news' visibility and spread (\sect\ref{sec:curation_amplification}); and the \textit{concentration of extremely biased and low factual content} (\sect\ref{sec:communities_not_users}), examining the distribution of links from the perspectives of who submitted them and what community they were submitted to.

We show that communities on \reddit exist across the left-right political spectrum, as measured by MBFC, but 74\% are ideologically center left.
We find that the diversity of left-leaning communities' membership is similar to that of equivalently right-leaning communities, but right-leaning communities have 105\% more politically varied news sources, as their members individually post more varied links.
This variance comes from the presence of links that are different from the community average, and in right-leaning communities, 74\% of such links are to relatively-more biased news sources, 35\% more than in left-leaning communities (\sect\ref{sec:communities_not_equal}).

We demonstrate that, regardless of the political leaning of the community,  community members' voting and crossposting (re-sharing) behavior reduces the impact of extremely biased and low factual news sources. Links to these news sources receive 20\% fewer upvotes (\sect\ref{sec:score}) and 30\% fewer exposures from crossposts compared to more neutral and higher factual content (\sect\ref{sec:crossposts}).
Furthermore, we find that users who submit such content leave \reddit 68\% more quickly than others (\sect\ref{sec:lifespan}).
These findings suggest that low factual content spreads more slowly and is amplified less on \reddit than has been reported for Twitter \cite{vosoughi2018spread, Bovet2019InfluenceOF}, \new{although we do not directly compare behavior across the two platforms. Differences between \reddit and Twitter} may stem from \reddit's explicit division into communities, or users' ability to downvote content, both of which help control content exposure.

Extremely biased and low factual content can be challenging to manage, as it is spread through many users, news sources, and communities.
We find that extremely biased and low factual content is spread by an even broader set of users and communities relative to news content as a whole, exacerbating this challenge (\sect\ref{sec:communities_not_users}). However, we find that 99\% of extremely biased or low factual content is still concentrated in 0.5\% of communities, lending credence to recent interventions 
at the community level~\cite{chandrasekharan2017cant_stay_here, chandrasekharan2020quarantined, saleem2018aftermath, ribiero2020migration}.

Our work demonstrates that additional research on news sharing online is especially needed on the topics of why users depart platforms and where they go, why false news appears to spread more quickly on Twitter than on \reddit, and how curation and amplification practices can manage influxes of extremely biased and low factual content.

Finally, we make all of our data and analyses publicly available\footnote{\papersite} to encourage future work on this important topic.

\section{Related Work}\label{sec:related}
\xhdr{Misinformation and Deceptive News}
Social news platforms have seen a continued increase in use and a simultaneous increase in concern regarding biased news and misinformation \cite{mitchell19pew,marwick2017media}. %
Recent studies have used network spread \cite{vosoughi2018spread, ferrara2017contagion, Bovet2019InfluenceOF}, content consumer \cite{allen2020evaluating}, and content producer \cite{linvill2020troll} approaches to assess the spread of misinformation.
In this work, we examine news sharing behavior from news sources who publish content with varied degrees of bias or factualness, building on related work that has analyzed social news based on the characteristics of a new source's audience~\cite{samory2020characterizing} or the type of content posted~\cite{glenski2018propagation}.

\xhdr{Polarization and Political Bias}
Many papers have recently been published on detecting political bias of online content either automatically \cite{baly2020detect, Demszky2019AnalyzingPI} or manually \cite{Ganguly2020EmpiricalEO, bozarth2020higher}. Others have examined bias in moderation of content, as opposed to biased content or news sources themselves \cite{Jiang2019BiasMT, Jiang2020ReasoningAP}. Echo chambers are a major consideration in understanding polarization, with papers focusing on their development \cite{Allison2020CommunalQA} and the role of news sources in echo chambers \cite{Horne2019DifferentSO}. Others have examined who shares what content with what political bias, but did so using implicit community structure~\cite{Samory2020CharacterizingTS}. In this work, we examine thousands of explicit communities on \reddit, characterizing their polarization by examining the political diversity of news sources shared within, and the diversity of the community members who contribute.

\xhdr{Moderation and Governance}
A large body of work has examined the role of moderation interventions such explanations \cite{Jhaver2019DoesTI}, content removal \cite{Chandrasekharan2018TheIH}, community bans \cite{chandrasekharan2017cant_stay_here, chandrasekharan2020quarantined, saleem2018aftermath} on outcomes such as migration \cite{ribiero2020migration}, harassment \cite{Matias2019PreventingHA} and harmful language use~\cite{Wadden2021Moderation}. Others have focused on moderators themselves \cite{matias2019civic_labor, Dosono2019ModerationPA}, and technological tools to assist them \cite{Jhaver2019HumanMachineCF, Zhang2020PolicyKitBG, Chandrasekharan2019CrossmodAC}, as well as self-moderation through voting \cite{glenski2017consumers, Risch2020TopCO} and community norms \cite{Fiesler2018RedditRC}. In contrast, our work informs the viability of different moderation strategies, specifically by examining the sharing and visibility of news content across thousands of communities. 
\section{Dataset \& Validation}\label{sec:method}
We analyze all \reddit submissions to extract links, and annotate links to news sources with their political bias and factualness using ratings from Media Bias/Fact Check.

\subsection{Reddit Content}\label{sec:reddit_data}
\reddit is the sixth most visited website in the world, and is widely studied due to its size, diversity of communities, and the public availability of its content~\cite{Medvedev2018TheAO}. 
Users can submit links or text (known as ``selfposts'') to specific communities, known as ``subreddits.''
Users may view submissions for a single community, or create a ``front page'' which aggregates submissions from all communities the user ``subscribes'' to.
Here, we focus on submissions over comments, as submissions are the primary mechanism for sharing content on \reddit, and users spend most of their time engaging with submissions \cite{glenski2017consumers}.

To create our dataset, we downloaded all public \reddit submissions from Pushshift \cite{baumgartner2020pushshift} posted between January 2015 and August 2019\footnote{August 2019 was the most recent month available at the time of this study.}, inclusive, for a total of 56 months 
of content (580 million submissions, 35 million unique authors, 3.4 million unique subreddits). For each submission, we extract the URLs of each linked-to website, which resulted in 559 million links\footnote{While link submissions by definition contain exactly one link, text submissions (selfposts) can include 0 or more links.}.
Additional summary statistics are included in Appendix~\ref{app:summary}

\xhdr{Ethical Considerations}
We value and respect the privacy and agency of all people potentially impacted by this work. 
All \reddit content analyzed in this study is publicly accessible, and Pushshift, from which we source our \reddit content, permits any user to request removal of their submissions at any time.
We take specific steps to protect the privacy of people included in our study \cite{Fiesler2018ParticipantPO}: we do not identify specific users, and we exclusively analyze data and report our results in aggregate. All analysis of data in this study was conducted in accordance with the Institutional Review Board at \new{ the University of Washington under ID \texttt{STUDY00011457}}.

\begin{figure}[t]
    \centering
    \includegraphics[width=1\columnwidth]{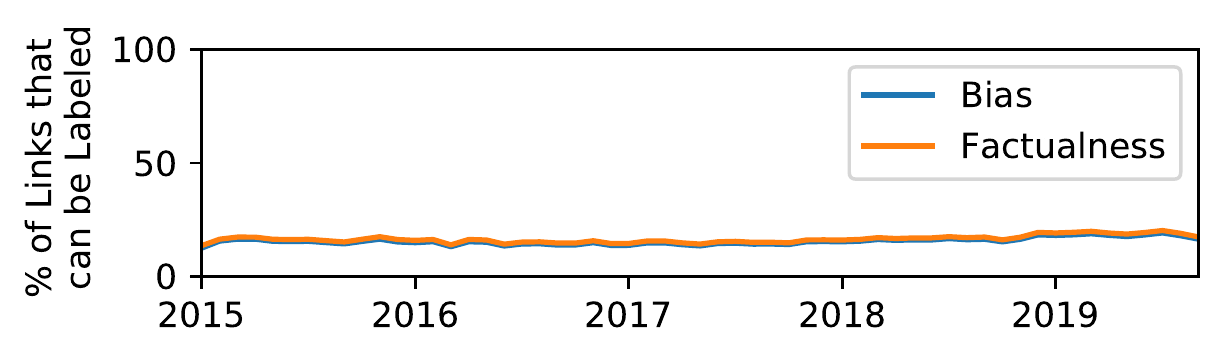}
    \vspace{-1em}
    \caption{The percentage of links that can be annotated using the MBFC labels is very consistent ($\pm$ 3.3\%) over time, suggesting that comparisons over time are not significantly impacted by changes in annotation coverage.}
    \label{fig:coverage}
\end{figure}

\subsection{Annotation of Links' News Sources}\label{sec:mbfc_labeling}
To identify and annotate links to news sources, we make use of Media Bias/Fact Check (hereafter MBFC), an independently run news source rating service. \citet{bozarth2020higher} find that ``the choice of traditional news lists [for fact checking] seems to not matter,'' when comparing 5 different news lists including MBFC. Therefore, we selected MBFC as it offers the largest set of labels of any news source rating service \cite{bozarth2020higher}.
MBFC provides ratings of the political bias (left to right) and factualness (low to high) of news outlets around the world, along with additional details and justifications for ratings, using a rigorous public methodology\footnote{\texttt{mediabiasfactcheck.com/methodology/}}. MBFC is widely used for labelling bias and factualness of news sources for downstream analysis \cite{heydari2019youtube, main2018alt_right, Starbird2017ExaminingTA, Darwish2017TrumpVH, nelimarkka2018social_media} and as ground truth for prediction tasks \cite{dinkov2019predicting, stefanov2020predicting}.

From MBFC's public reports on each news source, we extract the name of the news source, its website, and the political bias and factualness ratings. Bias is measured on a 7-point scale of `extreme left,' `left,' `center left,' `center,' `center right,' `right,' and `extreme right,' and is reported for 2,440 news sources. Factualness is measured on a 6-point scale of `very low factual,' `low factual,' `mixed factual,' `mostly factual,' `high factual,' and `very high factual,' and is reported for 2,676 news sources (as of April 2020).
For brevity, in the following analyses, we occasionally use the term `left leaning' to indicate a news source with a bias rating of `extreme left,' `left,' or `center left,' and the term `right leaning' to indicate a news source with a bias rating of `center right,' `right,' or `extreme right.'

\begin{figure}[t]
    \centering
    \includegraphics[width=\columnwidth]{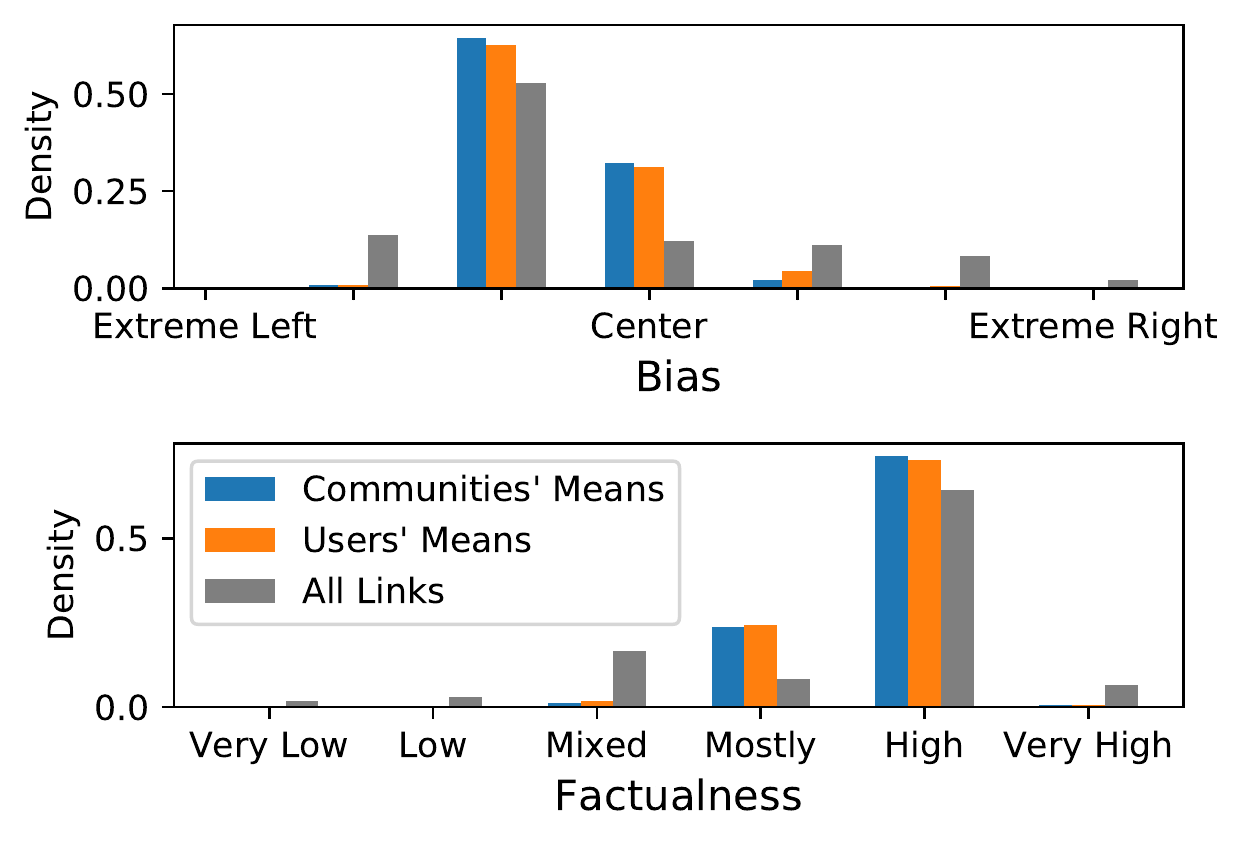}
    \vspace{-2em}
    \caption{Distributions of mean bias and factualness are quite similar for both the user and community units of analysis. Grey bars show the normalized total counts of links of each type across all of \reddit.}
    \label{fig:distrib_of_means}
\end{figure}

We then annotate the links extracted from \reddit submissions with the MBFC ratings using regular expressions to match the URL of the link with the domain of the corresponding news source. For example, a link to \texttt{www.rt.com/news/covid/} would be matched with the \texttt{rt.com} domain of RT, the Russian-funded television network, and annotated with a bias of `center right' and a factualness of `very low.'
\new{Links to URL shorteners such as \texttt{bit.ly} were excluded from labeling.}
We find that links to center left and high factual news sources are most common, accounting for 53\% and 64\% of all news links, respectively. Extreme left news source links are much less common, with 22.2 extreme right links for every 1 extreme left link (\fig~\ref{fig:distrib_of_means}).

\xhdr{Validation of MBFC Annotations} 
The use of fact checking sources such as MBFC is common practice for large scale studies, and MBFC in particular is widely used~\cite{dinkov2019predicting, stefanov2020predicting, heydari2019youtube, main2018alt_right, Starbird2017ExaminingTA, Darwish2017TrumpVH, nelimarkka2018social_media}. Additional confidence in MFBC annotations comes from the results of \citet{bozarth2020higher}, who find that (1) MBFC offers the largest set of biased and low factual news sources when compared among 5 fact checking datasets, and (2) the selection of a specific fact checking source has little impact on the evaluation of online content. Furthermore, we find that the coverage (the percentage of links that can be annotated using MBFC, excluding links to obvious non-news sources such as links to elsewhere on \reddit, to shopping sites, \etc)  is very consistent ($\pm$ 3.3\%) over the 4 year span of our dataset (\fig\ref{fig:coverage}). Additionally, \citet{bozarth2020higher} find that it is very rare for a news source's bias or factualness to change over time, suggesting that the potential `drift' of ratings over time should not affect our results.

\new{
\xhdr{Robustness Checks with Different Set of Annotations}
}
Lastly, we use an additional fact checking dataset from \citet{volkova2017separating}, consisting of 251 `verified' news sources and 210 `suspicious' news sources, as an additional point of comparison for validation. While the exact classes in the Volkova \etal dataset are not directly comparable to MBFC, we can create a comparable class by comparing links with a MBFC factualness rating of `mostly factual' or higher with Volkova \etal's `verified' news sources. In this case, when considering links that can be annotated using both datasets, MBFC and Volkova \etal have a Cohen's kappa coefficient of 0.82, indicating ``almost perfect'' inter-rater reliability \cite{Landis1977TheMO}. We examined if these differences could have an impact of downstream analysis and found this to be unlikely. For example, results computed separately using MBFC and Volkova \etal agree with one another with a Pearson's correlation of 0.98 on the task of identifying the number of `mostly factual' or higher links posted to a community.

\xhdr{Computing Mean Bias/Factualness}
As described above, MBFC labels for bias and factualness are ordinal, yet for many analyses, it is useful to have numeric labels (\eg computing the variance of links in a community). To convert from MBFC's categorical labels to a numeric scale, we use a mapping of (-3, -2, -1, 0, 1, 2, 3) to assign `extreme left' links a numeric bias value of -3, `left' links a value of -2, `center left' links a value of -1, `center' links a value of 0, and positive values to map to the equivalent categories on the right.
While this choice is somewhat arbitrary, it is consistent with the linear spacing between bias levels given by MBFC. Furthermore, we explored different mappings, including nonlinear ones, and found that our results are robust to different mappings. As such, we use the mapping given above as it is easiest to interpret.
We use a similar mapping of (0, 1, 2, 3, 4, 5) to assign `very low factual' links a numeric value of 0, `low factual' links a value of 1, \etc, with `very high factualness' links assigned a value of 5.

These numeric values are used to compute users' and communities' \textit{mean bias} and \textit{mean factualness}, central constructs in our analyses. To do so, we simply take the average of the numeric bias and factualness values of the links by each user or in each community. 
For many of our analyses, we group users by rounding their mean bias/factualness to the nearest integer. Thus, when we describe a user as having a `left center bias,' we are indicating that the mean bias of the links they submitted is between -1.5 and -0.5.

The distributions of means are very similar for users and communities, with both closely following the overall distribution of news links on \reddit, shown with grey bars (\fig\ref{fig:distrib_of_means}). 
 74\% of communities and 73\% of users have a mean bias of approximately center left, and 65\% of communities and 62\% of users have a mean factualness of `high factual' (among users/communities with more than 10 links).

Similarly, we define \textit{user variance of bias} as the variance of the bias values of the links submitted by a user, and similarly \textit{community variance of bias} is defined as the variance of of the bias values of links submitted to a community. As with mean bias, we find that the distributions of user and community variance of bias are very similar to one another. The median user has a variance of 0.85, approximately the variance of a user with center bias who submits 62\% center links, 22\% center-left or center-right links, and 16\% left or right links. The median community has a variance of 0.91, approximately that of a community where 62\% of the content submitted has center bias, 20\% of the content has center-left or center-right bias, and 18\% of the content has left or right bias. Of course, a substantial amount of a community's variance comes from the variance of its userbase. We explore sources of this variance in \sect\ref{sec:communities_not_equal}.

\subsection{Estimating Potential Exposures to Content}
Links on \reddit do not have equal impact; some links are viewed by far more people than others. To understand the impact of certain types of content, we would like to understand how broadly that content is viewed. As view counts are not publicly available, we use the number of subscribers to the community that a link was posted to as an estimate for the number of \textit{potential exposures} to community members that this content may have had. While some users, \new{especially those without accounts}, view content from communities they are not subscribed to, subscription counts capture both active contributors and passive consumers within the community, which motivated our use of this proxy over other alternatives, \new{such as the number of votes}. 

As communities are constantly growing, we define the number of potential exposures to a link as the number of subscribers to the community the link was posted to \textit{at the time it was posted}. To estimate historic subscriber counts, we make use of archived Wayback machine snapshots of subreddit about pages, which provide the number of subscribers at the time of the snapshot.  These snapshots are available for the $\sim$3,500 largest subreddits. In addition, we collected the current (as of Dec. 29, 2020) subscriber count for the 25,000 largest subreddits, as well as the date the subreddit was created (at which point it had 0 subscribers). We use the present subscriber count, archived subscriber counts (if available), and the creation date, and linearly interpolate between these data points to create a historical estimate of the subscriber counts over time for each of the 25,000 largest (by number of posts) subreddits in our dataset. The resulting set of subscriber count data, when joined with our set of \reddit content, provides potential exposure estimates for 93.8\% of submissions. For the remaining 6.2\% of submissions, we intentionally, conservatively \emph{overestimate} the potential exposures by using the first percentile value (4 subscribers) from our subscriber count data. The effect of this imputation on our results is very minor as these only occur in communities with extremely little activity.

\section{Diversity of News within Communities}\label{sec:communities_not_equal}

In this section, we examine the factors that contribute to a community's variance of bias. This variance can come from a combination of two sources: (1) community members who are individually ideologically diverse (\textit{user diversity}), and (2) a diverse group of users with different mean biases (\textit{group diversity}).
High user diversity corresponds to a community whose members have high user variance (\eg users who are ideologically diverse individually), and high group diversity corresponds to a community with high variance of its members' mean bias (\eg a diverse group of users who may be ideologically consistent individually). Of course, these sources of variance are not mutually exclusive; \textit{overall community variance} is maximized when both user diversity \textit{and} group diversity are large.

\xhdr{Method}
This intuition can be formalized using the Law of Total Variance, which states that total community variance is exactly the sum of User Diversity (within-user variance) and Group Diversity (between-user variance):
\begin{align*}
    \var(\mathcal{B}_c) = \E[\var(\mathcal{B}_c|\mathcal{U})] + \var(\E[\mathcal{B}_c|\mathcal{U}])
\end{align*}

\noindent
where $\mathcal{B}_c$ is a random variable representing the bias of a link submitted to community $c$, and $\mathcal{U}$ is a random variable representing the user who submitted the link.

We compute user diversity and group diversity for each community. User diversity is given by taking the mean of each user's variance of bias, weighted by the number of labeled bias links that user submitted. Group diversity is given by taking the variance of each community members' mean user bias, again weighted by their number of labeled links. We then sum the user and group diversity values to compute the overall community variance of political bias.

To understand \textit{how} communities vary relative to their mean, we compute the balance of links in the adjacent relatively more- and less- biased categories. For example, a community with `left' mean bias has two adjacent categories: `extreme left' and `center left,' with `extreme left' being the relatively-more biased category, and `center left' being the relatively-less biased category.

\begin{figure}[t]
    \centering
    \includegraphics[trim={.1cm 0 .1cm 0cm},clip,width=\columnwidth]{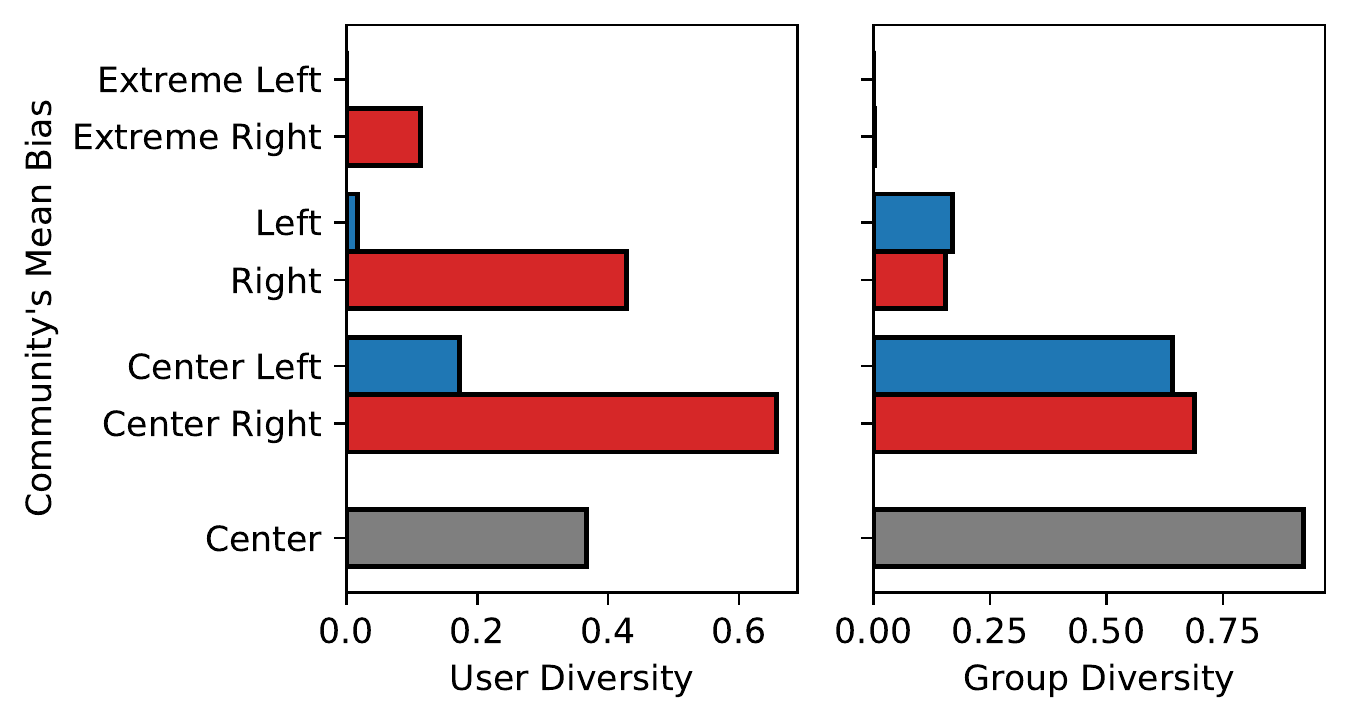}
    \vspace{-2em}
    \caption{While group diversity is similar between left- and right-leaning communities with a similar degree of bias (right panel), right-leaning communities have higher user diversity than equivalently biased communities on the left (left panel). As a result, right-leaning communities have higher overall variance around their community mean. Right-leaning communities also favor relatively-more biased links, when compared to left-leaning communities.}
    \label{fig:variance}
\end{figure}

\xhdr{Results}
Across all of \reddit, we find most (82\%) communities' group diversity constitutes a majority of their overall variance of bias.
When binned by their mean bias, we find that communities with extreme bias have, on average, lower total variance than communities closer to the middle of the spectrum (\fig\ref{fig:variance}). A community with mean bias of `extreme left' would be expected to have a lower total variance as there are no links with bias further left than `extreme left.' To control for this dynamic, we only compare symmetric labels: `extreme left' to `extreme right,' `left' to `right,' and `center left' to `center right.'

We find that right- and left-leaning communities have similar group diversity (\fig\ref{fig:variance}, right), but right-leaning communities (red) have 341\% more user diversity than equivalently left-leaning communities, on average (\fig\ref{fig:variance}, left). As a result, the average overall variance is 105\% greater for right-leaning communities than left-leaning communities.
Interestingly, we find that a larger share of right-leaning communities' variance is in more biased categories, relative to the community mean.
74\% of right-leaning communities' adjacent links are relatively-more biased, compared to 55\% for left-leaning communities, in other words, an increase of 35\% $\left(\frac{74\%}{55\%}\right)$.

\xhdr{Implications}
These results suggest that members of communities on the left and right have comparable group diversity, indicating the range of users are equally similar to one another. However, right-leaning communities have higher user diversity, indicating that the individual users themselves tend to submit links to news sources with a larger variety of political leaning. This creates higher overall variance of political bias in right-leaning communities, however these right-leaning communities also contain more links with higher bias, relative to the community mean, as opposed to more relatively-neutral news sources.

\section{Impact of Current Curation and Amplification Behaviors}\label{sec:curation_amplification}
The impact of content on \reddit is affected by users' behavior: how long they stay on the platform, how they vote, and how they amplify. In this section, we examine user longevity and turnover, community acceptance of biased and low factual content, and amplification through crossposting.

\begin{figure}[t]
    \centering
    \includegraphics[width=\columnwidth]{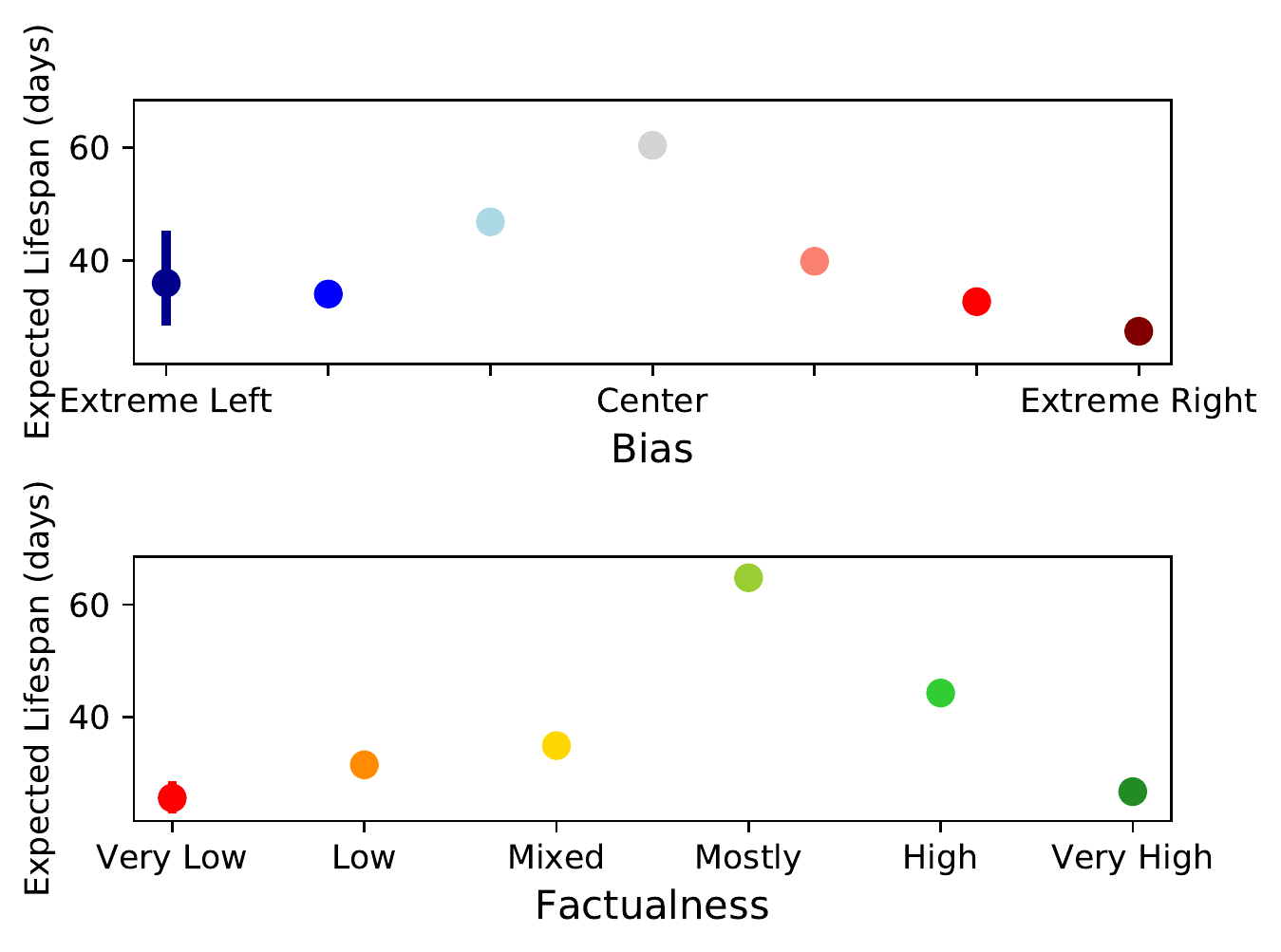}
    \vspace{-2em}
    \caption{
    Users with extreme mean bias stay on \reddit less than half as long as users with center mean bias. Users with low and very low mean factualness also leave more quickly, but expected lifespan decreases as users' mean factualness increases past `mixed factual'.
    Across all figures, error bars correspond to bootstrapped 95\% confidence intervals (and may be too small to be visible).
    }
    \label{fig:lifespan}
\end{figure}

\subsection{User Lifespan}\label{sec:lifespan}
Do users who post extremely biased or low factual content stay on \reddit as long as other users?

\xhdr{Method}
We compute each user's lifespan on the platform by measuring how long they stay active on the platform after their first submission. We define ``active'' as posting at least once every 30 days, as in \citet{waller2019generalists}. We group users by their mean bias and factualness, and for each group, compute the expected lifespan of the group members.

\xhdr{Results}
We find that expected lifespan is longer for users who typically submit less politically biased content, with users whose mean bias is near center remaining on \reddit for approximately twice as long as users with extreme or moderate mean bias, on average (\fig\ref{fig:lifespan}, top). This result holds regardless of whether users are left- or right-leaning. Users with a mean factualness close to `mixed factual' or lower leave \reddit 68\% faster than users whose mean factualness is near `mostly factual' (\fig\ref{fig:lifespan}, bottom). However, we also find that users' expected lifespan decreases dramatically as their mean factualness increases to `high' or `very high' levels of factualness.

\xhdr{Implications}
These results suggest that users who mostly post links to extremely biased or low factual news sources leave \reddit more quickly than other users. We can only speculate as to the causes of this faster turnover, but we note that users who stay on \reddit the longest tend to post links to the types of news sources that are most prevalent (grey bars in \fig\ref{fig:distrib_of_means} show overall prevalence of each type of link).

The faster turnover suggests that users sharing this type of content leave relatively early, limiting their impact on their communities. However, faster turnover also may make user-level interventions such as bans less effective, as these sanctions have shorter-lived impact when the users they are made against leave the site more quickly. Future research could examine why users leave, whether they rejoin with new accounts in violation of \reddit policy, and the efficacy of restrictions of new accounts.

\begin{figure*}[t]
    \centering
    \includegraphics[width=\textwidth]{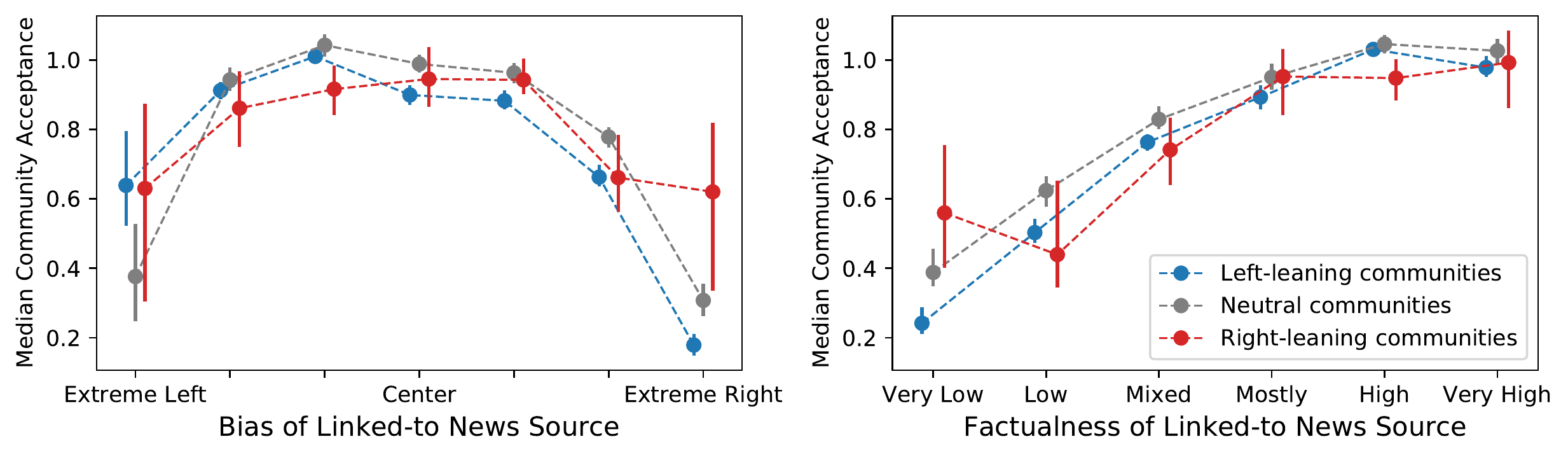}
    \vspace{-2em}
    \caption{Regardless of the political leaning of the community, extremely biased content is less accepted by communities than content closer to center. Similarly, low and very low factual content is less accepted than higher factual content. Points perturbed on the x-axis to aid readability.}
    \label{fig:score}
\end{figure*}

\subsection{Acceptance of Biased or Low Factual Content}\label{sec:score}
How do communities respond to politically biased or low factual content?

\xhdr{Method}
On \reddit, community members curate content in their communities by voting submissions up or down, which affects its position on the community feed~\cite{glenski2017consumers}. A submission's `score' is defined by \reddit as approximately the number of upvotes minus the number of downvotes that post receives. 
The score has been used in previous work as a proxy for a link's reception by a community \cite{waller2019generalists, Datta2019ExtractingIC}.
Links submitted to larger communities are seen by more users and therefore receive more votes. Therefore, we normalize each link's score by dividing by the mean score of all submissions in that community; links with a normalized score over $1$ are more accepted than average, and links with a score under $1$ are less accepted than average.
In accordance with \reddit's ranking algorithm, submissions with higher normalized score appear higher in the feed viewed by community members, and stay in this position for longer \cite{Medvedev2018TheAO}.

To compute the \textit{community acceptance} of links of a given bias or factualness, we average the normalized score of all links of that type in that community. We then take the median community acceptance across all left-leaning, right-leaning, and neutral communities. Here we use the median as it is more resilient to outliers than the mean.

\xhdr{Results}
We find that, regardless of the community's political leaning, median expected community acceptance is 18\% lower for extremely biased content than other content (\fig~\ref{fig:score}). For left-leaning and neutral communities, community acceptance decreases monotonically as factualness drops below `high.' However, we observe that right leaning communities are 167\% ($p=0.0002$) more accepting of extreme right biased and 85\% ($p=0.004$) more accepting of very low factual content than left-leaning and neutral communities (Mann--Whitney $U$ significance tests).

\xhdr{Implications}
This suggests that across \reddit, communities are sensitive to extremely biased and low factual content, and users' voting behavior is fairly effective at reducing the acceptance of this content. However, curation does not seem to result in better-than-average acceptance for any content---no median acceptance values are significantly ($p<0.05$) above 1, as non-news content tends to receive higher community acceptance than news content.

Previous research has found that on Twitter, news that failed fact-checking spread more quickly and was seen more widely than news that passed a fact-check \cite{vosoughi2018spread}. Interestingly, we find evidence that behavior on \reddit is somewhat different, with median left-leaning, right-leaning, and neutral communities all being less accepting of low and very low factual content.
\new{Importantly, our methodology differs from \citet{vosoughi2018spread} in that we use bias and factualness evaluations that were applied to  entire news sources, as opposed to the fact checking of specific news articles, limiting direct comparisons. Furthermore, we do not analyze the time between an initial post and its subsequent amplification, and so cannot directly comment on the `speed' of amplification. We do find evidence, however, that highly biased content on \reddit is less upvoted than more neutral content.}

\new{These difference may in part be explained by differences between reddit's and Twitter's mechanisms for impacting the visibility of content.}
Whereas Twitter users are only able to \new{increase visibility by retweeting, liking, replying to, or quoting} content, on \reddit, users may downvote to decrease visibility of content they object to.
We speculate that this may partially explain the differences in acceptance that we find between \reddit and Twitter.

\subsection{Selective Amplification of News Content }\label{sec:crossposts}
How does %
amplification of content affect exposure to biased and low factual content? On \reddit, users are not only able to submit links to \textit{external} content (such as news sites), but users are also able to submit links to \textit{internal} content elsewhere on \reddit, effectively re-sharing and therefore \emph{amplifying} content by increasing its visibility on the site. %
This is commonly known as `crossposting,' and often occurs when a user submits a post from one subreddit to another subreddit, although such re-sharing of internal content can happen within a single community as well. Here, we seek to understand the role that amplification through crossposts has on \reddit user's exposure to various kinds of content.

\xhdr{Method}
To identify the political bias and factualness of crossposted content, we identify all crossposted links to news sources, and propagate the label of the crossposted link. Then, we compute the fraction of total potential exposures from crossposts for each bias/factualness category.

\xhdr{Results}
We find that amplification via crossposting has an overall small effect on the potential exposures of news content. While 10\% of all news links are crossposts, only 1\% of potential exposures to news links are due to crossposts. This suggests that the majority of crossposts are content posted in
relatively larger communities re-shared to relatively smaller communities with relatively fewer subscribers,
diminishing the impact of amplification via crossposting. As such, \textit{direct} links to news sites have a far greater bearing on \reddit users' exposure to news content than crossposts.

\begin{figure}[t]
    \centering
    \includegraphics[width=\columnwidth]{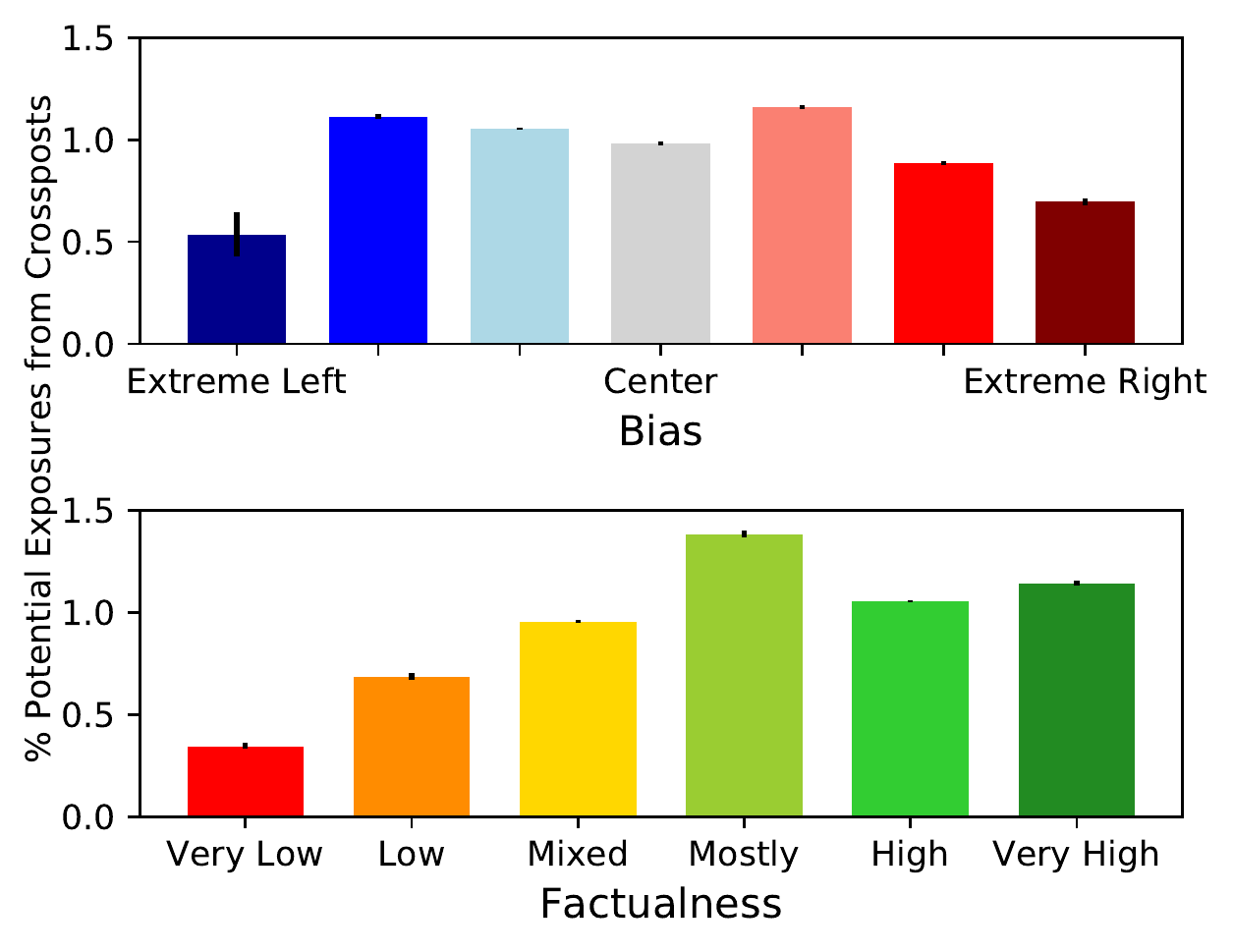}
    \vspace{-2em}
    \caption{ Extremely biased and low factual content is amplified by crossposts relatively less than other content. Regardless of the bias or factualness of the content, while crossposts are responsible for more than 750 billion potential exposures, they make up only 1\% of total potential exposures, suggesting that direct links to news sources play an especially important role in content distribution.
    }
    
    \label{fig:amplification}
\end{figure}

However, the role of crossposts in exposing users to new content is still important, as crossposts account for more than 750 billion potential exposures. We find that extremely biased and low factual content is amplified less than other content, as shown in \fig\ref{fig:amplification}, which illustrates the percentage of total potential exposures that come from crossposts for each bias/factualness category. \reddit users exposed to center left biased, center biased, or center right biased content are 53\% more likely to be exposed to this content via amplification than \reddit users exposed to extremely biased content. Similarly, \reddit users exposed to `mostly factual' or higher factualness content are 217\% more likely to be exposed to such content via amplification than \reddit users exposed to very low factual content.

\xhdr{Implications}
Given that only 1\% of potential exposures are from amplifications, understanding the way that \textit{direct} links to external content are shared is critical to understanding the sharing of news content on \reddit more broadly.

The relative lower amplification of extremely biased and very low factual content suggests users' sensitivity to the bias and factualness of the content they are re-sharing. As in \sect\ref{sec:score}, this suggests differences between \reddit and Twitter, where content that failed a fact-check has been found to spread more quickly than fact-checked content~\cite{vosoughi2018spread}. We speculate that this may be due to structural differences between the two platforms. On \reddit, users primarily consume content through subscriptions to communities, not other users. This may explain the diminished impact of re-sharing on \reddit compared to Twitter.

\section{Concentrations of Extremely Biased or Low Factual News Content}\label{sec:communities_not_users}
It is critical to understand where different news content is concentrated in order to best inform strategies for monitoring and managing its spread online. In this section, we examine how extremely biased and low factual content is distributed across users, communities, and news sources. We also compare the concentration of extremely biased and low factual content to all content.

\xhdr{Method}
We consider three types of content: (1) news content with extreme bias or low factualness, (2) all news content, and (3) all content (including non-news).
We group each of these types of content by three perspectives: the user who posted the content, the community it was posted to, and the news source (or domain, in the case of all content) linked to. We then take the cumulative sum of potential exposures across the users, communities, and news sources, to compute the fraction of potential exposures contributed by the top $n$\% of users, communities, and news sources.
We  repeat this process, replacing the number of potential exposures with the total number of links, to consider the concentration of links being submitted, regardless of visibility.

\begin{figure}[t]
    \centering
    \includegraphics[trim={.1cm 0 .1cm 0cm},clip,width=\columnwidth]{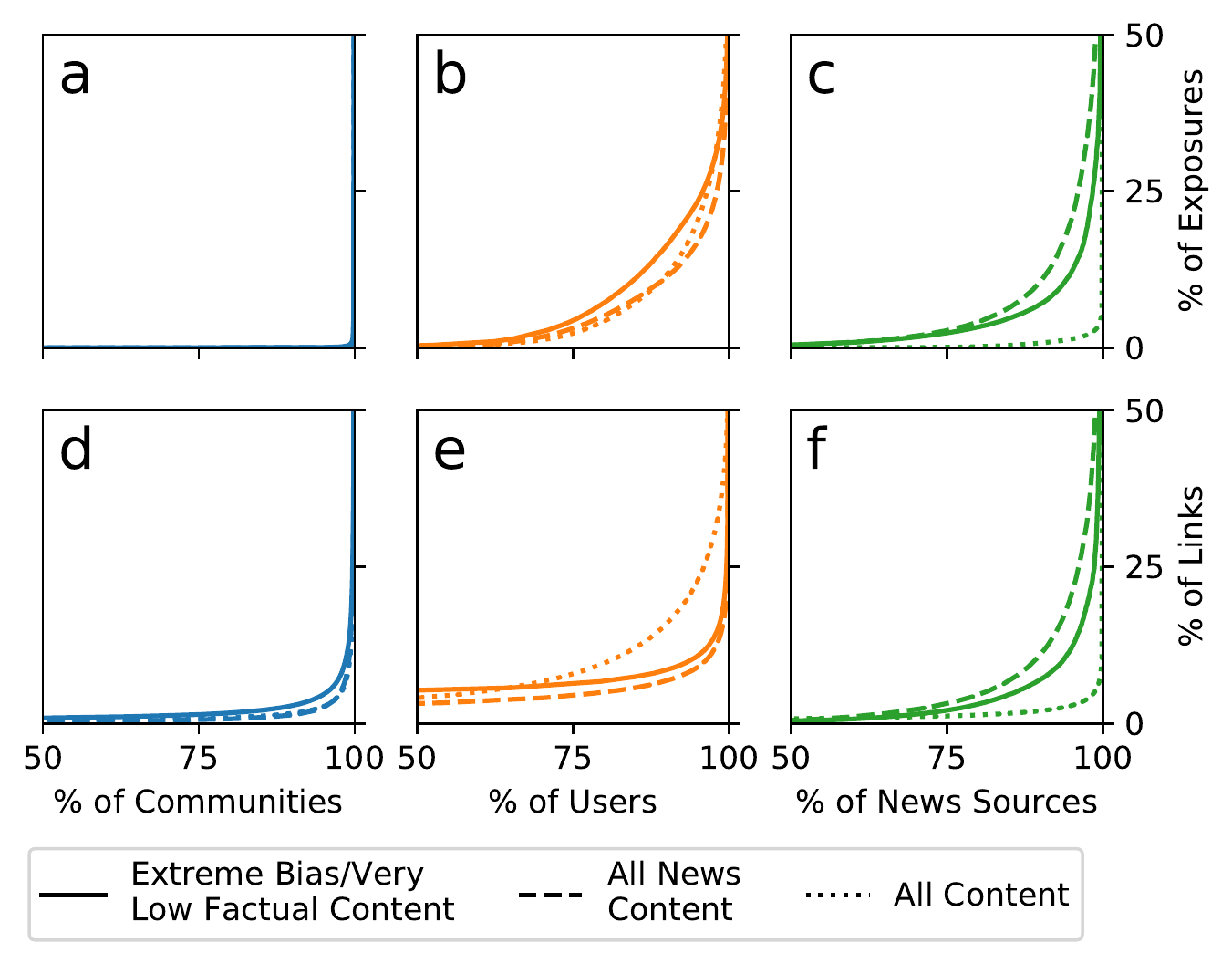}
    \vspace{-1em}
    \caption{
    When compared to all content on \reddit (dotted line), extremely biased or low factual content (solid line) is more broadly distributed, making it harder to detect, regardless of the community, user, or news source perspective. However, 99\% of potential exposures to extremely biased or low factual content are restricted to only 0.5\% of communities. Here, a curve closer to the lower-right corner indicates a more extreme concentration. Note that axis limits do not extend from 0 to 100\%.}
    \label{fig:lorenz}
\end{figure}

\xhdr{Results}
We find that overall, extremely biased and low factual content is highly concentrated across all three perspectives, but is especially concentrated in a small number of communities, where 99\% of potential exposures stem from a mere 109 (0.5\%) communities (Gini coefficient=0.997) (\fig\ref{fig:lorenz}a).
No matter the perspective, exposures to extremely biased or low factual content (solid line) are less concentrated than all content (dotted line) (\fig\ref{fig:lorenz}abc).

Under the community and news source perspectives, exposures (\fig\ref{fig:lorenz}ac) are more concentrated than links (\fig\ref{fig:lorenz}df). While links are already concentrated in a small share of communities, some communities are especially large, and therefore content from these communities receives a disproportionate share of potential exposures. This is not the case for users, as the distributions of exposures (\fig\ref{fig:lorenz}b) are less concentrated than the distributions of links (\fig\ref{fig:lorenz}e). This indicates that while some users submit a disproportionate share of links, these are not the users whose links receive the largest potential exposure, as potential exposure is primarily a function of submitting links to large communities.

\xhdr{Implications}
The extreme concentration of extremely biased or low factual content amongst a tiny fraction of communities supports \reddit's recent and high profile decision to take sanctions against entire communities, not just specific users \cite{nyt2021reddit_ban}. These decisions have been extensively studied~\cite{chandrasekharan2017cant_stay_here, chandrasekharan2020quarantined, thomas2021behavior, saleem2018aftermath, ribiero2020migration}. 
While this content is relatively less concentrated amongst users, in absolute terms, this content is still fairly concentrated, with 10\% of users contributing 84\% of potential exposures. As such, moderation sanctions against users can still be effective \cite{matias2019civic_labor}. \new{We note that the concentration of extremely biased or low factual content amongst a small fraction of users is similar to what has been found on Twitter \cite{Grinberg2019FakeNO}, although methodological differences preclude a direct comparison.}

\section{Discussion}\label{sec:discussion}
\xhdr{Summary \& Implications}
In this work, we analyze all 580 million submissions to \reddit from 2015-2019, and annotate 35 million links to news sources with their political bias and factualness using Media Bias/Fact Check. We find:
\begin{itemize}[leftmargin=*]
\item Right-leaning communities' links to news sources have 105\% greater variance in their political bias than left-leaning communities. When right-leaning communities link to news sources that are different than the community average, they link to relatively-more biased sources 35\% more often than left-leaning communities (\sect\ref{sec:communities_not_equal}).
\item Existing curation and amplification behaviors moderately reduce the impact of highly biased and low factual content. This suggests that \reddit \new{differs somewhat from Twitter}, perhaps due to its explicit community structure, or the ability for users to downvote content (\sect\ref{sec:curation_amplification}). 
\item Highly biased and low factual content tends to be shared by a broader set of users and in a broader set of communities than news content as a whole. Furthermore, the distribution of this content is more concentrated in a small number of communities than a small number of users, as 99\% of exposures to extremely biased or low factual content stem from only 0.5\% or 109 communities (\sect\ref{sec:communities_not_users}). This lends credence to recent \reddit interventions at the community level, including bans and quarantines.
\end{itemize}

\xhdr{Limitations} 
One limitation of our analyses is the use of a single news source rating service, MBFC. However, the selection of news source rating annotation sets has been found to have a minimal impact on research results \cite{bozarth2020higher}. MBFC is the largest (that we know of) dataset of news sources' bias and factualness, and is widely used~\cite{dinkov2019predicting, stefanov2020predicting, heydari2019youtube, Starbird2017ExaminingTA, Darwish2017TrumpVH}. 
More robust approaches could combine annotations from multiple sources, and we find that MBFC annotations agree with the \citet{volkova2017separating} dataset with a Pearson Correlation of 0.96 on an example downstream task (\sect\ref{sec:mbfc_labeling}).

Our focus is on the bias and factualness of news sources shared online. We do not consider factors such as the content of links (\eg shared images, specific details of news stories), or the context in which links are shared (\eg sentiment of a submission's comments). These factors are important areas for future work, and are outside the scope of this paper.

While MBFC (and by extension, our annotations) includes news sources from around the world, our analyses, especially the left-right political spectrum and associated colors, takes a US-centric approach.
Polarization and misinformation are challenges across the globe \cite{reuters_news_report_2020}, and more work is needed on other cultural contexts.

Our paper explores the impact of curation and amplification practices, but not the impact of community moderators who are a critical component of \reddit's moderation pipeline \cite{matias2019civic_labor}. Future work could examine news content removed by moderators.

\new{
Finally, we are limited by the unavailability of data on which users view what content. While we use subreddits' subscriber counts to estimate exposures to content, more granular data would enable us to better understand the impact of shared news articles, for example, the percentage of users who are exposed to extremely biased or low factual content \cite{Grinberg2019FakeNO}.
}

\section{Conclusion}\label{sec:conclusion}
Biased and inaccurate news shared online are significant problems, with real harms across our society. Large-scale studies of news sharing online are critical for understanding the scale and dynamics of these problems. We presented the largest study to date of news sharing behavior on \reddit, and found that right-leaning communities have more politically varied and relatively-more biased links than left-leaning communities, current voting and re-sharing behaviors are moderately effective at reducing the impact of extremely biased and low factual content, and that such content is extremely concentrated in a small number of communities.
We make our dataset of news sharing on \reddit public, \new{in order to support further research}\footnote{\papersite}.

\section*{Acknowledgements}
This research was supported by the Laboratory Directed Research and Development Program at Pacific Northwest National Laboratory, a multiprogram national laboratory operated by Battelle for the U.S. Department of Energy. This research was supported by the Office for Naval Research, NSF grant IIS-1901386,  the Bill \& Melinda Gates Foundation (INV-004841), and a Microsoft AI for Accessibility grant. 

\bibliography{bibliography.bib}

\begin{thebibliography}{62}
\providecommand{\natexlab}[1]{#1}
\providecommand{\url}[1]{\texttt{#1}}
\providecommand{\urlprefix}{URL }
\expandafter\ifx\csname urlstyle\endcsname\relax
  \providecommand{\doi}[1]{doi:\discretionary{}{}{}#1}\else
  \providecommand{\doi}{doi:\discretionary{}{}{}\begingroup
  \urlstyle{rm}\Url}\fi

\bibitem[{Allen et~al.(2020)Allen, Howland, Mobius, Rothschild, and
  Watts}]{allen2020evaluating}
Allen, J.; Howland, B.; Mobius, M.; Rothschild, D.; and Watts, D.~J. 2020.
\newblock Evaluating the fake news problem at the scale of the information
  ecosystem.
\newblock \emph{Science Advances} 6(14).

\bibitem[{Allison and Bussey(2020)}]{Allison2020CommunalQA}
Allison, K.; and Bussey, K. 2020.
\newblock Communal Quirks and Circlejerks: A Taxonomy of Processes Contributing
  to Insularity in Online Communities.
\newblock In \emph{ICWSM}.

\bibitem[{Baly et~al.(2020)Baly, Da~San~Martino, Glass, and
  Nakov}]{baly2020detect}
Baly, R.; Da~San~Martino, G.; Glass, J.; and Nakov, P. 2020.
\newblock We Can Detect Your Bias: Predicting the Political Ideology of News
  Articles.
\newblock In \emph{EMNLP}, 4982--4991. ACL.
\newblock \doi{10.18653/v1/2020.emnlp-main.404}.
\newblock \urlprefix\url{https://www.aclweb.org/anthology/2020.emnlp-main.404}.

\bibitem[{Baumgartner et~al.(2020)Baumgartner, Zannettou, Keegan, Squire, and
  Blackburn}]{baumgartner2020pushshift}
Baumgartner, J.; Zannettou, S.; Keegan, B.; Squire, M.; and Blackburn, J. 2020.
\newblock The Pushshift Reddit Dataset.
\newblock In \emph{ICWSM}.

\bibitem[{Bovet and Makse(2019)}]{Bovet2019InfluenceOF}
Bovet, A.; and Makse, H. 2019.
\newblock Influence of fake news in Twitter during the 2016 US presidential
  election.
\newblock \emph{Nature Communications} 10.

\bibitem[{Bozarth, Saraf, and Budak(2020)}]{bozarth2020higher}
Bozarth, L.; Saraf, A.; and Budak, C. 2020.
\newblock Higher Ground? How Groundtruth Labeling Impacts Our Understanding of
  Fake News about the 2016 U.S. Presidential Nominees.
\newblock In \emph{ICWSM}.

\bibitem[{Chandrasekharan et~al.(2019)Chandrasekharan, Gandhi, Mustelier, and
  Gilbert}]{Chandrasekharan2019CrossmodAC}
Chandrasekharan, E.; Gandhi, C.; Mustelier, M.~W.; and Gilbert, E. 2019.
\newblock Crossmod: A Cross-Community Learning-based System to Assist Reddit
  Moderators.
\newblock \emph{CHI} 3: 1 -- 30.

\bibitem[{Chandrasekharan et~al.(2020)Chandrasekharan, Jhaver, Bruckman, and
  Gilbert}]{chandrasekharan2020quarantined}
Chandrasekharan, E.; Jhaver, S.; Bruckman, A.; and Gilbert, E. 2020.
\newblock Quarantined! Examining the Effects of a Community-Wide Moderation
  Intervention on Reddit.
\newblock \emph{ArXiv} abs/2009.11483.

\bibitem[{Chandrasekharan et~al.(2017)Chandrasekharan, Pavalanathan,
  Srinivasan, Glynn, Eisenstein, and
  Gilbert}]{chandrasekharan2017cant_stay_here}
Chandrasekharan, E.; Pavalanathan, U.; Srinivasan, A.; Glynn, A.; Eisenstein,
  J.; and Gilbert, E. 2017.
\newblock You Can't Stay Here: The Efficacy of Reddit's 2015 Ban Examined
  Through Hate Speech.
\newblock \emph{CHI} 1: 31:1--31:22.

\bibitem[{Chandrasekharan et~al.(2018)Chandrasekharan, Samory, Jhaver, Charvat,
  Bruckman, Lampe, Eisenstein, and Gilbert}]{Chandrasekharan2018TheIH}
Chandrasekharan, E.; Samory, M.; Jhaver, S.; Charvat, H.; Bruckman, A.; Lampe,
  C.; Eisenstein, J.; and Gilbert, E. 2018.
\newblock The Internet's Hidden Rules.
\newblock \emph{CHI} 2: 1 -- 25.

\bibitem[{Choi et~al.(2020)Choi, Chun, Oh, Han et~al.}]{choi2020rumor}
Choi, D.; Chun, S.; Oh, H.; Han, J.; et~al. 2020.
\newblock Rumor propagation is amplified by echo chambers in social media.
\newblock \emph{Scientific Reports} 10(1): 1--10.

\bibitem[{Darwish, Magdy, and Zanouda(2017)}]{Darwish2017TrumpVH}
Darwish, K.; Magdy, W.; and Zanouda, T. 2017.
\newblock Trump vs. Hillary: What Went Viral During the 2016 US Presidential
  Election.
\newblock In \emph{SocInfo}.

\bibitem[{Datta and Adar(2019)}]{Datta2019ExtractingIC}
Datta, S.; and Adar, E. 2019.
\newblock Extracting Inter-community Conflicts in Reddit.
\newblock In \emph{ICWSM}.

\bibitem[{Demszky et~al.(2019)Demszky, Garg, Voigt, Zou, Gentzkow, Shapiro, and
  Jurafsky}]{Demszky2019AnalyzingPI}
Demszky, D.; Garg, N.; Voigt, R.; Zou, J.; Gentzkow, M.; Shapiro, J.; and
  Jurafsky, D. 2019.
\newblock Analyzing Polarization in Social Media: Method and Application to
  Tweets on 21 Mass Shootings.
\newblock In \emph{NAACL-HLT}.

\bibitem[{Dinkov et~al.(2019)Dinkov, Ali, Koychev, and
  Nakov}]{dinkov2019predicting}
Dinkov, Y.; Ali, A.; Koychev, I.; and Nakov, P. 2019.
\newblock Predicting the Leading Political Ideology of YouTube Channels Using
  Acoustic, Textual, and Metadata Information.
\newblock In \emph{INTERSPEECH}.

\bibitem[{Dosono and Semaan(2019)}]{Dosono2019ModerationPA}
Dosono, B.; and Semaan, B.~C. 2019.
\newblock Moderation Practices as Emotional Labor in Sustaining Online
  Communities: The Case of AAPI Identity Work on Reddit.
\newblock \emph{CHI} .

\bibitem[{Ferrara(2017)}]{ferrara2017contagion}
Ferrara, E. 2017.
\newblock Contagion dynamics of extremist propaganda in social networks.
\newblock \emph{Information Sciences} 418: 1--12.

\bibitem[{Fiesler et~al.(2018)Fiesler, Jiang, McCann, Frye, and
  Brubaker}]{Fiesler2018RedditRC}
Fiesler, C.; Jiang, J.~A.; McCann, J.; Frye, K.; and Brubaker, J.~R. 2018.
\newblock Reddit Rules! Characterizing an Ecosystem of Governance.
\newblock In \emph{ICWSM}.

\bibitem[{Fiesler and Proferes(2018)}]{Fiesler2018ParticipantPO}
Fiesler, C.; and Proferes, N. 2018.
\newblock “Participant” Perceptions of Twitter Research Ethics.
\newblock \emph{Social Media + Society} 4.

\bibitem[{Ganguly et~al.(2020)Ganguly, Kulshrestha, An, and
  Kwak}]{Ganguly2020EmpiricalEO}
Ganguly, S.; Kulshrestha, J.; An, J.; and Kwak, H. 2020.
\newblock Empirical Evaluation of Three Common Assumptions in Building
  Political Media Bias Datasets.
\newblock In \emph{ICWSM}.

\bibitem[{Geiger(2019)}]{geiger_2019_online_news}
Geiger, A. 2019.
\newblock Key findings about the online news landscape in America.
\newblock
  \urlprefix\url{https://www.pewresearch.org/fact-tank/2019/09/11/key-findings-about-the-online-news-landscape-in-america/}.

\bibitem[{{Glenski}, {Pennycuff}, and {Weninger}(2017)}]{glenski2017consumers}
{Glenski}, M.; {Pennycuff}, C.; and {Weninger}, T. 2017.
\newblock Consumers and Curators: Browsing and Voting Patterns on Reddit.
\newblock \emph{IEEE TCSS} 4(4): 196--206.
\newblock \doi{10.1109/TCSS.2017.2742242}.

\bibitem[{Glenski, Weninger, and Volkova(2018)}]{glenski2018propagation}
Glenski, M.; Weninger, T.; and Volkova, S. 2018.
\newblock Propagation from deceptive news sources who shares, how much, how
  evenly, and how quickly?
\newblock \emph{IEEE TCSS} 5(4): 1071--1082.

\bibitem[{Grinberg et~al.(2019)Grinberg, Joseph, Friedland, Swire-Thompson, and
  Lazer}]{Grinberg2019FakeNO}
Grinberg, N.; Joseph, K.; Friedland, L.; Swire-Thompson, B.; and Lazer, D.
  2019.
\newblock Fake news on Twitter during the 2016 U.S. presidential election.
\newblock \emph{Science} 363: 374 -- 378.

\bibitem[{Heydari et~al.(2019)Heydari, Zhang, Appel, Wu, and
  Ranade}]{heydari2019youtube}
Heydari, A.; Zhang, J.; Appel, S.; Wu, X.; and Ranade, G. 2019.
\newblock YouTube Chatter: Understanding Online Comments Discourse on
  Misinformative and Political YouTube Videos.

\bibitem[{Horne, N{\o}rregaard, and Adali(2019)}]{Horne2019DifferentSO}
Horne, B.; N{\o}rregaard, J.; and Adali, S. 2019.
\newblock Different Spirals of Sameness: A Study of Content Sharing in
  Mainstream and Alternative Media.
\newblock In \emph{ICWSM}.

\bibitem[{Isaac and Conger(2021)}]{nyt2021reddit_ban}
Isaac, M.; and Conger, K. 2021.
\newblock Reddit bans forum dedicated to supporting Trump.
\newblock \emph{The New York Times}
  \urlprefix\url{https://www.nytimes.com/2021/01/08/us/politics/reddit-bans-forum-dedicated-to-supporting-trump-and-twitter-permanently-suspends-his-allies-who-spread-conspiracy-theories.html}.

\bibitem[{Jhaver et~al.(2019)Jhaver, Birman, Gilbert, and
  Bruckman}]{Jhaver2019HumanMachineCF}
Jhaver, S.; Birman, I.; Gilbert, E.; and Bruckman, A. 2019.
\newblock Human-Machine Collaboration for Content Regulation.
\newblock \emph{TOCHI} 26: 1 -- 35.

\bibitem[{Jhaver, Bruckman, and Gilbert(2019)}]{Jhaver2019DoesTI}
Jhaver, S.; Bruckman, A.; and Gilbert, E. 2019.
\newblock Does Transparency in Moderation Really Matter?
\newblock \emph{CHI} 3: 1 -- 27.

\bibitem[{Jiang, Robertson, and Wilson(2019)}]{Jiang2019BiasMT}
Jiang, S.; Robertson, R.~E.; and Wilson, C. 2019.
\newblock Bias Misperceived: The Role of Partisanship and Misinformation in
  YouTube Comment Moderation.
\newblock In \emph{ICWSM}.

\bibitem[{Jiang, Robertson, and Wilson(2020)}]{Jiang2020ReasoningAP}
Jiang, S.; Robertson, R.~E.; and Wilson, C. 2020.
\newblock Reasoning about Political Bias in Content Moderation.
\newblock In \emph{AAAI}.

\bibitem[{Kharratzadeh and {\"U}stebay(2017)}]{Kharratzadeh2017USPE}
Kharratzadeh, M.; and {\"U}stebay, D. 2017.
\newblock US Presidential Election: What Engaged People on Facebook.
\newblock In \emph{ICWSM}.

\bibitem[{Kouzy et~al.(2020)Kouzy, Jaoude, Kraitem, Alam, Karam, Adib, Zarka,
  Traboulsi, Akl, and Baddour}]{Kouzy2020CoronavirusGV}
Kouzy, R.; Jaoude, J.~A.; Kraitem, A.; Alam, M. B.~E.; Karam, B.; Adib, E.;
  Zarka, J.; Traboulsi, C.; Akl, E.~A.; and Baddour, K. 2020.
\newblock Coronavirus Goes Viral: Quantifying the COVID-19 Misinformation
  Epidemic on Twitter.
\newblock \emph{Cureus} 12.

\bibitem[{Landis and Koch(1977)}]{Landis1977TheMO}
Landis, J.; and Koch, G. 1977.
\newblock The measurement of observer agreement for categorical data.
\newblock \emph{Biometrics} 33 1: 159--74.

\bibitem[{Linvill and Warren(2020)}]{linvill2020troll}
Linvill, D.~L.; and Warren, P.~L. 2020.
\newblock Troll factories: Manufacturing specialized disinformation on Twitter.
\newblock \emph{Political Communication} 1--21.

\bibitem[{Main(2018)}]{main2018alt_right}
Main, T.~J. 2018.
\newblock \emph{The Rise of the Alt-Right}.
\newblock Brookings Institution Press.

\bibitem[{Marwick and Lewis(2017)}]{marwick2017media}
Marwick, A.; and Lewis, R. 2017.
\newblock Media manipulation and disinformation online.
\newblock \emph{Data \& Society}
  \url{https://datasociety.net/library/media-manipulation-and-disinfo-online/}.

\bibitem[{Matias(2019{\natexlab{a}})}]{Matias2019PreventingHA}
Matias, J. 2019{\natexlab{a}}.
\newblock Preventing harassment and increasing group participation through
  social norms in 2,190 online science discussions.
\newblock \emph{PNAS} 116: 9785 -- 9789.

\bibitem[{Matias(2019{\natexlab{b}})}]{matias2019civic_labor}
Matias, J.~N. 2019{\natexlab{b}}.
\newblock The Civic Labor of Volunteer Moderators Online.
\newblock \emph{Social Media + Society} 5(2): 2056305119836778.
\newblock \doi{10.1177/2056305119836778}.
\newblock \urlprefix\url{https://doi.org/10.1177/2056305119836778}.

\bibitem[{Medvedev, Lambiotte, and Delvenne(2018)}]{Medvedev2018TheAO}
Medvedev, A.; Lambiotte, R.; and Delvenne, J. 2018.
\newblock The anatomy of Reddit: An overview of academic research.
\newblock \emph{ArXiv} abs/1810.10881.

\bibitem[{Mitchell et~al.(2019)Mitchell, Gottfried, Srocking, Walker, and
  Fedeli}]{mitchell19pew}
Mitchell, A.; Gottfried, J.; Srocking, G.; Walker, M.; and Fedeli, S. 2019.
\newblock Many Americans Say Made-Up News Is a Critical Problem That Needs To
  Be Fixed.
\newblock \emph{Pew Research Center Science and Journalism}
  \url{https://www.journalism.org/2019/06/05/many-americans-say-made-up-news-is-a-critical-problem-that-needs-to-be-fixed/}.

\bibitem[{Nelimarkka, Laaksonen, and Semaan(2018)}]{nelimarkka2018social_media}
Nelimarkka, M.; Laaksonen, S.-M.; and Semaan, B. 2018.
\newblock Social Media Is Polarized, Social Media Is Polarized: Towards a New
  Design Agenda for Mitigating Polarization.
\newblock In \emph{ACM DIS}, 957--970. United States: ACM.
\newblock \doi{10.1145/3196709.3196764}.
\newblock ACM conference on Designing Interactive Systems, DIS ; Conference
  date: 09-06-2018 Through 13-06-2018.

\bibitem[{Newman(2020)}]{reuters_news_report_2020}
Newman, N. 2020.
\newblock \emph{Digital News Report}.
\newblock Reuters Institute.

\bibitem[{Qazvinian et~al.(2011)Qazvinian, Rosengren, Radev, and
  Mei}]{qazvinian2011rumor}
Qazvinian, V.; Rosengren, E.; Radev, D.; and Mei, Q. 2011.
\newblock Rumor has it: Identifying misinformation in microblogs.
\newblock In \emph{EMNLP}, 1589--1599.

\bibitem[{Rajadesingan, Resnick, and Budak(2020)}]{Rajadesingan2020QuickCL}
Rajadesingan, A.; Resnick, P.; and Budak, C. 2020.
\newblock Quick, Community-Specific Learning: How Distinctive Toxicity Norms
  Are Maintained in Political Subreddits.
\newblock In \emph{ICWSM}.

\bibitem[{Recuero, Soares, and Gruzd(2020)}]{Recuero2020HyperpartisanshipDA}
Recuero, R.; Soares, F.~B.; and Gruzd, A.~A. 2020.
\newblock Hyperpartisanship, Disinformation and Political Conversations on
  Twitter: The Brazilian Presidential Election of 2018.
\newblock In \emph{ICWSM}.

\bibitem[{Ribeiro et~al.(2020)Ribeiro, Jhaver, Zannettou, Blackburn,
  Cristofaro, Stringhini, and West}]{ribiero2020migration}
Ribeiro, M.~H.; Jhaver, S.; Zannettou, S.; Blackburn, J.; Cristofaro, E.~D.;
  Stringhini, G.; and West, R. 2020.
\newblock Does Platform Migration Compromise Content Moderation? Evidence from
  r/The\_Donald and r/Incels.
\newblock \emph{ArXiv} abs/2010.10397.

\bibitem[{Risch and Krestel(2020)}]{Risch2020TopCO}
Risch, J.; and Krestel, R. 2020.
\newblock Top Comment or Flop Comment? Predicting and Explaining User
  Engagement in Online News Discussions.
\newblock In \emph{ICWSM}.

\bibitem[{Saleem and Ruths(2018)}]{saleem2018aftermath}
Saleem, H.~M.; and Ruths, D. 2018.
\newblock The Aftermath of Disbanding an Online Hateful Community.
\newblock \emph{ArXiv} abs/1804.07354.

\bibitem[{Samory, Abnousi, and
  Mitra(2020{\natexlab{a}})}]{samory2020characterizing}
Samory, M.; Abnousi, V.~K.; and Mitra, T. 2020{\natexlab{a}}.
\newblock Characterizing the Social Media News Sphere through User Co-Sharing
  Practices.
\newblock In \emph{ICWSM}, volume~14, 602--613.

\bibitem[{Samory, Abnousi, and
  Mitra(2020{\natexlab{b}})}]{Samory2020CharacterizingTS}
Samory, M.; Abnousi, V.~K.; and Mitra, T. 2020{\natexlab{b}}.
\newblock Characterizing the Social Media News Sphere through User Co-Sharing
  Practices.
\newblock In \emph{ICWSM}.

\bibitem[{Shearer and Matsa(2018)}]{shearer_2018}
Shearer, E.; and Matsa, K.~E. 2018.
\newblock News Use Across Social Media Platforms 2018.
\newblock
  \urlprefix\url{https://www.journalism.org/2018/09/10/news-use-across-social-media-platforms-2018/}.

\bibitem[{Starbird(2017)}]{Starbird2017ExaminingTA}
Starbird, K. 2017.
\newblock Examining the Alternative Media Ecosystem Through the Production of
  Alternative Narratives of Mass Shooting Events on Twitter.
\newblock In \emph{ICWSM}.

\bibitem[{Stefanov et~al.(2020)Stefanov, Darwish, Atanasov, and
  Nakov}]{stefanov2020predicting}
Stefanov, P.; Darwish, K.; Atanasov, A.; and Nakov, P. 2020.
\newblock Predicting the Topical Stance and Political Leaning of Media using
  Tweets.
\newblock In \emph{ACL}, 527--537. Online: ACL.
\newblock \doi{10.18653/v1/2020.acl-main.50}.
\newblock \urlprefix\url{https://www.aclweb.org/anthology/2020.acl-main.50}.

\bibitem[{Tasnim, Hossain, and Mazumder(2020)}]{Tasnim2020ImpactOR}
Tasnim, S.; Hossain, M.; and Mazumder, H. 2020.
\newblock Impact of Rumors and Misinformation on COVID-19 in Social Media.
\newblock \emph{Journal of Preventive Medicine and Public Health} 53: 171 --
  174.

\bibitem[{Thomas et~al.(2021)Thomas, Riehm, Glenski, and
  Weninger}]{thomas2021behavior}
Thomas, P.~B.; Riehm, D.; Glenski, M.; and Weninger, T. 2021.
\newblock Behavior Change in Response to Subreddit Bans and External Events.

\bibitem[{Volkova et~al.(2017)Volkova, Shaffer, Jang, and
  Hodas}]{volkova2017separating}
Volkova, S.; Shaffer, K.; Jang, J.~Y.; and Hodas, N. 2017.
\newblock Separating Facts from Fiction: Linguistic Models to Classify
  Suspicious and Trusted News Posts on Twitter.
\newblock In \emph{ACL}, 647--653. Vancouver, Canada: ACL.
\newblock \doi{10.18653/v1/P17-2102}.
\newblock \urlprefix\url{https://www.aclweb.org/anthology/P17-2102}.

\bibitem[{Vosoughi, Mohsenvand, and Roy(2017)}]{vosoughi2017rumor}
Vosoughi, S.; Mohsenvand, M.~N.; and Roy, D. 2017.
\newblock Rumor gauge: Predicting the veracity of rumors on Twitter.
\newblock \emph{KDD} 11(4): 1--36.

\bibitem[{Vosoughi, Roy, and Aral(2018)}]{vosoughi2018spread}
Vosoughi, S.; Roy, D.; and Aral, S. 2018.
\newblock The spread of true and false news online.
\newblock \emph{Science} 359(6380): 1146--1151.
\newblock ISSN 0036-8075.
\newblock \doi{10.1126/science.aap9559}.
\newblock \urlprefix\url{https://science.sciencemag.org/content/359/6380/1146}.

\bibitem[{Wadden et~al.(2021)Wadden, August, Li, and
  Althoff}]{Wadden2021Moderation}
Wadden, D.; August, T.; Li, Q.; and Althoff, T. 2021.
\newblock The Effect of Moderation on Online Mental Health Conversations.
\newblock In \emph{ICWSM}.

\bibitem[{Waller and Anderson(2019)}]{waller2019generalists}
Waller, I.; and Anderson, A. 2019.
\newblock Generalists and Specialists: Using Community Embeddings to Quantify
  Activity Diversity in Online Platforms.
\newblock In \emph{The World Wide Web Conference}, WWW '19, 1954–1964. New
  York, NY, USA: Association for Computing Machinery.
\newblock ISBN 9781450366748.
\newblock \doi{10.1145/3308558.3313729}.
\newblock \urlprefix\url{https://doi.org/10.1145/3308558.3313729}.

\bibitem[{Zhang, Hugh, and Bernstein(2020)}]{Zhang2020PolicyKitBG}
Zhang, A.~X.; Hugh, G.; and Bernstein, M.~S. 2020.
\newblock PolicyKit: Building Governance in Online Communities.
\newblock \emph{UIST} \urlprefix\url{https://doi.org/10.1145/3379337.3415858}.

\end{thebibliography}

\appendix
\twocolumn
\section{Dataset Summary}\label{app:summary}

\new{
Our dataset was created from all public \reddit submissions posted between January 2016 and August 2019, the most recent data available at the time of this study. These submissions were downloaded using the Pushshift archives~\cite{baumgartner2020pushshift}, and consist of 580 million submissions, 35 million unique authors, and 3.4 million unique subreddits.
As each submission may consist of 0 or more links, the dataset includes a total of 559 million links. These links are to 5.1 million unique domains, of which we are able to label 2,801 unique domains with annotations from MBFC.
}

\new{
The following table shows the number of links in the dataset, as well as the number of unique news sources, for each bias category.
}

\vspace{1em}
\begin{tabular}{l|r|r}
\textbf{Bias} & \textbf{\# of Links} & \textbf{\# of News Sources} \\ \hline
Extreme Left  & 15,157               & 51                          \\
Left          & 3,023,382            & 364                         \\
Center Left   & 17,648,711           & 544                         \\
Center        & 4,494,687            & 442                         \\
Center Right  & 4,254,705            & 263                         \\
Right         & 3,226,828            & 352                         \\
Extreme Right & 997,703              & 423                         \\
\textit{Unlabeled} & \textit{525,443,378} & \textit{}              \\

\end{tabular}
\vspace{1em}

\new{
\noindent
The following table shows the number of links in the dataset, as well as the number of unique news sources, for each factualness category.
}

\vspace{1em}

\begin{tabular}{l|r|r}
\textbf{Factualness} & \textbf{\# of Links} & \textbf{\# of News Sources} \\ \hline
Very Low             & 609,229              & 72                          \\
Low                  & 749,202              & 369                         \\
Mixed                & 7,116,130            & 677                         \\
Mostly               & 2,217,719            & 110                         \\
High                 & 22,055,943           & 1,313                       \\
Very High            & 2,263,604            & 134                         \\
\textit{Unlabeled}   & \textit{524,092,724} & \textit{}                   \\
\end{tabular}
\vspace{1em}

\noindent
The dataset may be downloaded from our website at \papersite

\end{document}